%% file: D0mixing.tex
\documentclass[twocolumn,twoside,slac_two]{revtex4}
\usepackage{graphicx}
\usepackage{multirow}
\usepackage{colordvi}
\usepackage{graphicx}
\usepackage{dcolumn}
\usepackage{bm}
\usepackage{fancyhdr}
\input babarsym
\input note1706symbols

\def\half{\ensuremath{{\textstyle{1 \over 2}}}}
\def\pri{^{\, \prime }}

\def\ket#1{\left| #1\right\rangle}

\begin{document}

\title{\boldmath$\Dz$--\boldmath$\Dzb$ Mixing at \babar}

%

\author{A. Seiden}
\affiliation{Santa Cruz Institute for Particle Physics,
  University of California, Santa Cruz, CA, USA\\
  Representing the \babar\ Collaboration}

\begin{abstract}
The \babar\ and Belle collaborations have recently found evidence
for mixing within the $D$ meson system.  We present some of the mixing search
techniques used by \babar\
and their status as of the beginning of the summer 2007.  These have
culminated in a measurement in the $K\pi$ decay final state of the $D$ that is
inconsistent with the no-mixing hypothesis with a significance of 3.9 standard
deviations.  
\end{abstract}
\maketitle
\thispagestyle{fancy}


\section{Introduction}

Mixing among the lightest neutral mesons of each flavor has traditionally
provided important information on the electroweak interactions, the CKM matrix,
and the possible virtual constituents that can lead to mixing.  Among the
long-lived mesons, the $D$ meson system exhibits the smallest mixing phenomena. 
The B-factories have now accumulated sufficient luminosity to observe mixing in
the $D$ system and we can expect to see more detailed results as more luminosity is
accumulated and additional channels sensitive to mixing are analyzed.  The
B-factories produce about 1.3 million Charm events per fb$^{-1}$ of integrated
luminosity accumulated.  The \babar\ integrated luminosity of about 384 fb$^{-1}$ used
for the evidence for mixing result we will present corresponds to about 500
million charm events produced.  The present \babar\ integrated luminosity is
approximately 500 fb$^{-1}$.  \babar\ is a high acceptance general-purpose detector
providing excellent tracking, vertexing, particle ID, and neutrals detection. 
All of these capabilities are crucial for making the difficult mixing measurement.

\section{Mixing Measureables for the \boldmath$D$ System}

The propagation eigenstates, including the electroweak interactions
for the $D$ mesons are given by:
\begin{equation}
\ket{D_{1,2}} = p\ket{\Dz}\pm q\ket{\Dzb}, \quad |p|^2 + |q|^2 = 1 .
\label{eq:eigenstates}
\end{equation}
Propagation parameters that determine the time-evolution
for the two states are given by:
\begin{equation}
\overline\Gamma = \half(\Gamma_1 + \Gamma_2),\quad \Delta M = M_1 - M_2, \quad
  \Delta\Gamma = (\Gamma_1 - \Gamma_2); 
\label{eq:progparam}
\end{equation}
with the observable oscillations determined by the scaled parameters
\begin{equation}
x = \frac{\Delta M}{\overline\Gamma},\quad y = 
\frac{\Delta\Gamma}{2\overline\Gamma} .
\label{eq:scaledparam}
\end{equation}

In the case of CP conservation the two $D$ eigenstates are the CP even and odd
combinations.  We will choose $D_1$ to be the CP even state.  The sign choice for
the mass and width difference varies among papers, we  use the choice above.

Assuming CP conservation, small mixing parameters, and an initial state tagged as a $\Dz$, we can write the time dependence to first order in $x$ and $y$:
\begin{equation}
D(t) = \left(\Dz + \Dzb (-y-ix) \frac{\overline\Gamma}{2} t\right) 
       e^{-(\overline\Gamma/2+i\overline m)t}.
\label{eq:Tdependence}
\end{equation}
Projecting this onto a final state $f$ gives to first order the amplitude for
finding $f$:
\begin{equation}
\left(A_f + \bar A_f (-y-ix) \frac{\overline\Gamma}{2} t\right)
                e^{-(\overline\Gamma/2+i\overline m) t}.
\label{eq:famplitude}
\end{equation}

This leads to a number of ways to measure the effect of mixing, for example:

(1) Wrong sign semileptonic decays.  Here $A_f$ is zero and we measure directly
the quantity, after integrating over decay times:
\begin{equation}                                       
R_M = (x^2 + y^2)/2.
\label{eq:WSdecay}
\end{equation}

Limits using this measurement, however, are not yet sensitive enough to get
down to the $10^{-4}$ level for $R_M$.  Using 334 fb$^{-1}$ of data, electron
decays only, and a double tag technique, \babar\ measures $R_M = 0.4 \times
10^{-4}$,  with a 68\% confidence interval $(-5.6, 7.4)\times
10^{-4}$\cite{slac12494}.   

(2) Cabibbo favored, right sign (RS) hadronic decays (for example
$K^-\pi^+$).  These are used to measure the average lifetime, with the
correction from the term involving $x$ and $y$ usually ignored (provides a
correction of $O(10^{-3})$).

(3) Singly suppressed decays (for example $\KpKm$ or $\pipi$).  In this case
tagging the initial state isn't necessary.  For CP even final states: $A_f =\bar
A_f$.  This provides the most direct way to measure $y$.  With tagging we can
also check for CP violation, by looking at the value of $y$ for each tag type. 
\babar\ will be updating this measurement with the full statistics later this
year.  The initial \babar\ measurement was based on 91 fb$^{-1}$ and gave the
result $y = 0.8\%$, with statistical and systematic errors each about 0.4\%
\cite{aubert2003}, 
measurement \cite{staric2007}. 

4) Doubly suppressed and mixed, wrong sign (WS) decays (for example
$\Kp\pim$.  Mixing leads to an exponential term multiplied by both a
linear and a quadratic term in $t$. The quadratic term has a universal
form depending on $R_M$.  For any point in the decay phase space the decay
rate is given by
\begin{equation}
\left(|A_f|^2 + |A_f|\ |\bar A_f|\ y\pri \overline\Gamma t + |\bar A_f|^2
    R_M \frac{(\overline\Gamma t)^2}{2}\right) e^{-\overline\Gamma t}.
\label{eq:Tws}
\end{equation}

Here $y\pri  = y \cos\delta - x\sin\delta$, where $\delta$ is a strong
phase difference between the Cabibbo favored and doubly suppressed
amplitudes.  For the $\Kp\pim$ decay there is just the one phase and the
ratio of $|A_f|^2$ to $|\bar A_f|^2$ is defined to be $R_D$.  For
multibody decays the strong phase  varies over the phase space and the term
proportional to $t$ will involve a sum with different phases if we add all
events in a given channel.

\babar\ has analyzed the decay channel $\Kp\pim\piz$, with a mass cut that selects
mostly $\Kp\rho^-$ decays, the largest channel for the Cabibbo allowed
amplitude arising from mixing.  Based on 230 fb$^{-1}$, \babar\ measures 
\cite{aubert2006}
\begin{eqnarray}
&\alpha y\pri = (-1.2^{+0.6}_{-0.8} \pm 0.2)\%  \nonumber \\
&R_M = (0.023^{+0.18}_{-0.13}\pm 0.004)\%.
\end{eqnarray}

The parameter $\alpha$ allows for the phase variation over the region
summed over.  A fit to the full Dalitz plot would allow more events to be
used in the mixing study.  This, however, requires a model for all the
resonant and smooth components that contribute to the given channel, which
may introduce uncertainties.  \babar\  is working on such a fit, which
will be based on approximately 1500 signal events.

Another important 3-body channel is the $K_s\pipi$ decay channel. Analysis of
this channel was pioneered by CLEO \cite{asner2005}.  
 It
contains: CP-even, CP-odd, and mixed-CP resonances.  Now one must 
correctly model the
relative amounts of CP-odd and CP-even contributions (including smooth
components) to get the correct lifetime difference.  This channel also 
provides the possibility to directly measure $x$.  \babar\  is  working on
this channel; Belle has published their results \cite{zhang}. 

In the Standard Model $y$ and $x$ are mainly due to long-distance effects. 
They may be comparable in value but this depends on physics that is difficult
to model.  Long-distance effects control how complete the SU(3) cancellation
is, which would make both parameters vanish in the symmetry limit.  The exact
values therefore depends on SU(3) violations in matrix elements and phase
space.  Also, the sign of $x/y$ provides an important measurement.  One might
expect the $x$ and $y$ parameters to be in the range $O(10^{-3}$ to
$10^{-2}$).  Thus the present data are consistent with the Standard Model. 
Searches for CP violation are important goals of the B-factories, since
observation at a non-negligible level would signify new physics.  

We will turn now to the strongest evidence for $D$-mixing from \babar,
using the $K\pi$ final state.  This result has recently been published
\cite{aubert2007}.  

\section{Analysis of the \boldmath$K\pi$ channel}
We study the right-sign (RS), Cabibbo-favored (CF)
decay $\Dztokpi$~\cite{footnote} and the
wrong-sign (WS) decay $\DztokpiWS$.  The latter can be produced via
the doubly Cabibbo-suppressed (DCS) decay $\DztokpiWS$ 
or via mixing followed by a CF
decay $\Dz \rightarrow \Dzbtokpi$.
The DCS decay has a small rate $R_D$
of order $\tan^4\theta_{\rm C} \approx 0.3\%$ relative to the CF decay
with $\theta_{\rm C}$ the Cabibbo angle.
We tag the  \Dz at production using the 
decay $\Dstp\to\pisoft^+\Dz$ where the 
$\pisoft^+$ is referred to as the ``slow pion''. 
In RS decays the $\pisoft^+$ and kaon have opposite charges,
while in WS decays the charges are the same.
The time dependence of the WS decay rate
is used to separate the contributions of DCS decays from \Dz-\Dzb mixing.

We study both \CP-conserving and \CP-violating cases.
For the \CP-conserving case, we fit for the parameters $R_D$, \xPrimeSq, and \yPrime. To search for \CP violation,
we apply Eq.~(\ref{eq:Tws}) 
to the \Dz and \Dzb samples separately, fitting for the parameters
\{$R_D^\pm$, $\xPrimePmSq$, $\yPrimePm$\} for \Dz($+$) decays and 
\Dzb($-$) decays.

\par
We select \Dz candidates by pairing  
oppositely-charged tracks with a $\Kmp\pipm$ invariant
mass \mKpi between $1.81$ and $1.92\gevcc$.
We require the $\pisoft^+$ to have a momentum in the laboratory frame
greater than $0.1\gevc$ and in the \epem center-of-mass (CM) frame
below $0.45\gevc$.
\par
To obtain the proper decay time \t\ and its error~\terr\ for each
\Dz candidate, we refit the \Kmp and \pipm tracks, constraining
them to originate from a 
common vertex. We also require the \Dz and $\pisoft^+$ 
to originate from a common vertex, 
constrained by the position and size of the \epem interaction region. The vertical RMS size of 
each beam is typically $6\mum$.
We require the \chisq\ probability of the vertex-constrained combined fit 
$\Pchisq$ to be at least $0.1\%$, and the $\mKpipis-\mKpi$ mass difference \dm to satisfy
$0.14<\dm<0.16\gevcc$.
\par
To remove \Dz candidates from $B$-meson decays and to reduce
combinatorial backgrounds, we require
each \Dz to have a momentum in the CM frame greater than
$2.5\gevc$. We require $-2 < \t<4\ps$
and $\terr<0.5\ps$
(the most probable value of \terr for signal events is $0.16\ps$).
For \Dstp\ candidates sharing one or more tracks with other \Dstp\ candidates,
we retain only the candidate with the highest \Pchisq.
After applying all criteria, we keep approximately 
1,229,000~RS and 64,000~WS
\Dz and \Dzb candidates.

The mixing parameters are determined in an unbinned, extended
maximum-likelihood fit to the RS and WS data samples over the four
observables \mKpi, \dm, $t$, and \terr. The fit is performed in several
stages. First, RS and WS signal and
background shape parameters are determined from a fit to \mKpi and
\dm, and are not varied in subsequent fits.  Next, the
\Dz proper-time resolution function and lifetime  are
determined in a fit to the RS data using \mKpi and \dm to separate the
signal and background components. We fit to the WS data sample using
three different models. The first model assumes both \CP conservation
and the absence of mixing.
The second model
allows for mixing, but assumes no
\CP violation.
The third model allows for both mixing and \CP violation.

The RS and WS \mdm\ distributions are described by four components:
signal, random $\pisoft^+$, misreconstructed \Dz and combinatorial
background. The signal component has a characteristic peak in both \mKpi and \dm.
The random $\pisoft^+$ component models reconstructed \Dz decays
combined with a random slow pion and has the same shape
in \mKpi as signal events, but does not peak in \dm.  Misreconstructed
\Dz events have one or more of the \Dz decay products either not
reconstructed or reconstructed with the wrong particle
hypothesis. They peak in \dm, but not in \mKpi. For RS events, most of
these are semileptonic \Dz\ decays. 
For WS events, the main contribution
is RS $\Dz\to\Kmpip$ decays 
where the
\Km and the \pip are misidentified as \pim and \Kp, respectively. 
Combinatorial background events are those not described by the
above components; they do not exhibit any peaking structure in
\mKpi or \dm.
\par
The functional forms of the probability density functions (PDFs) for
the signal and background components are chosen based on studies of
Monte Carlo (MC) samples. However, all parameters are determined from
two-dimensional likelihood fits to data over the full
\mKpi\ and \dm\ region. 

\begin{figure}[phtb]
  \centering
  \centerline{%
    \includegraphics[width=0.5\linewidth, clip=]{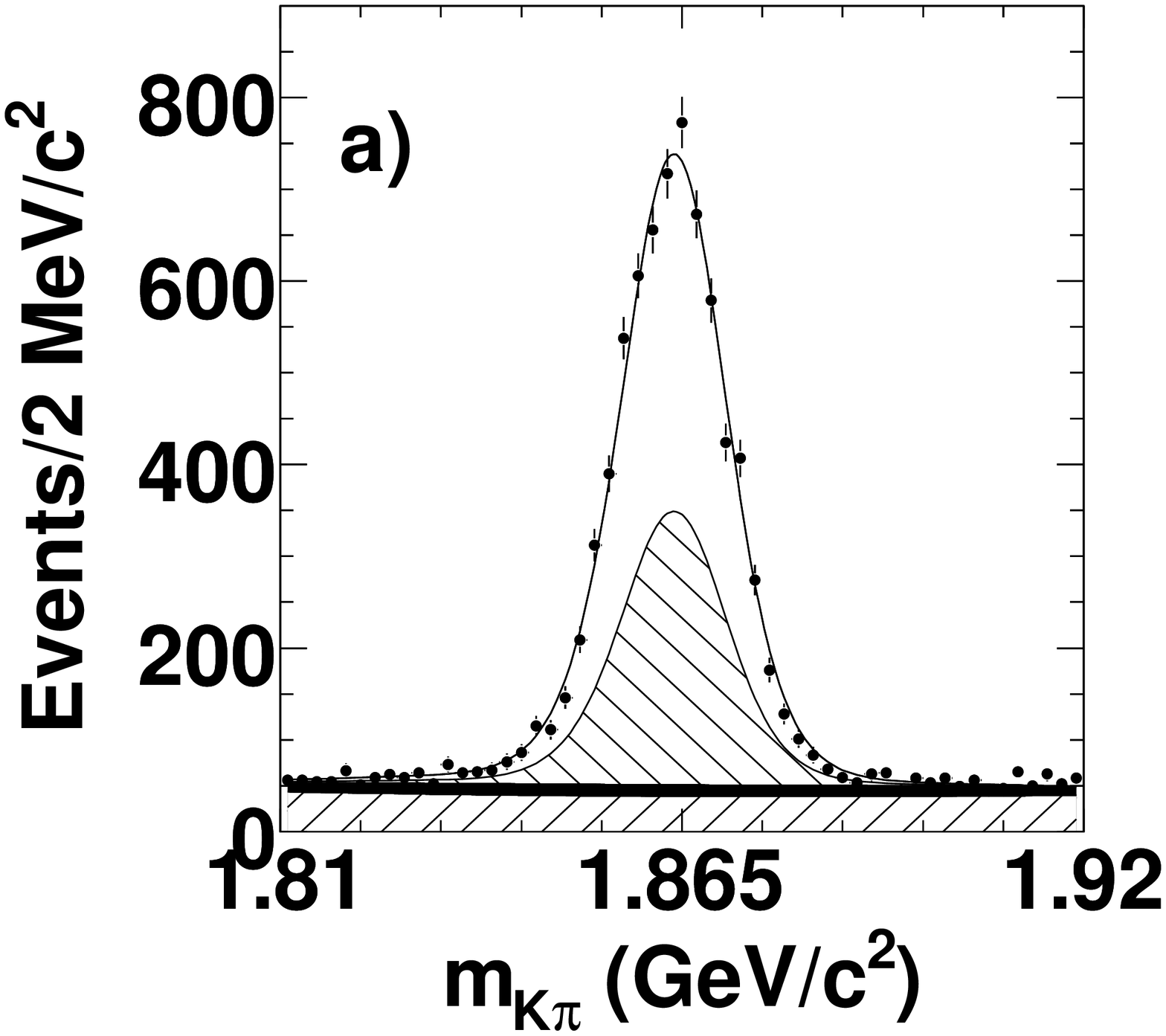}
    \includegraphics[width=0.5\linewidth, clip=]{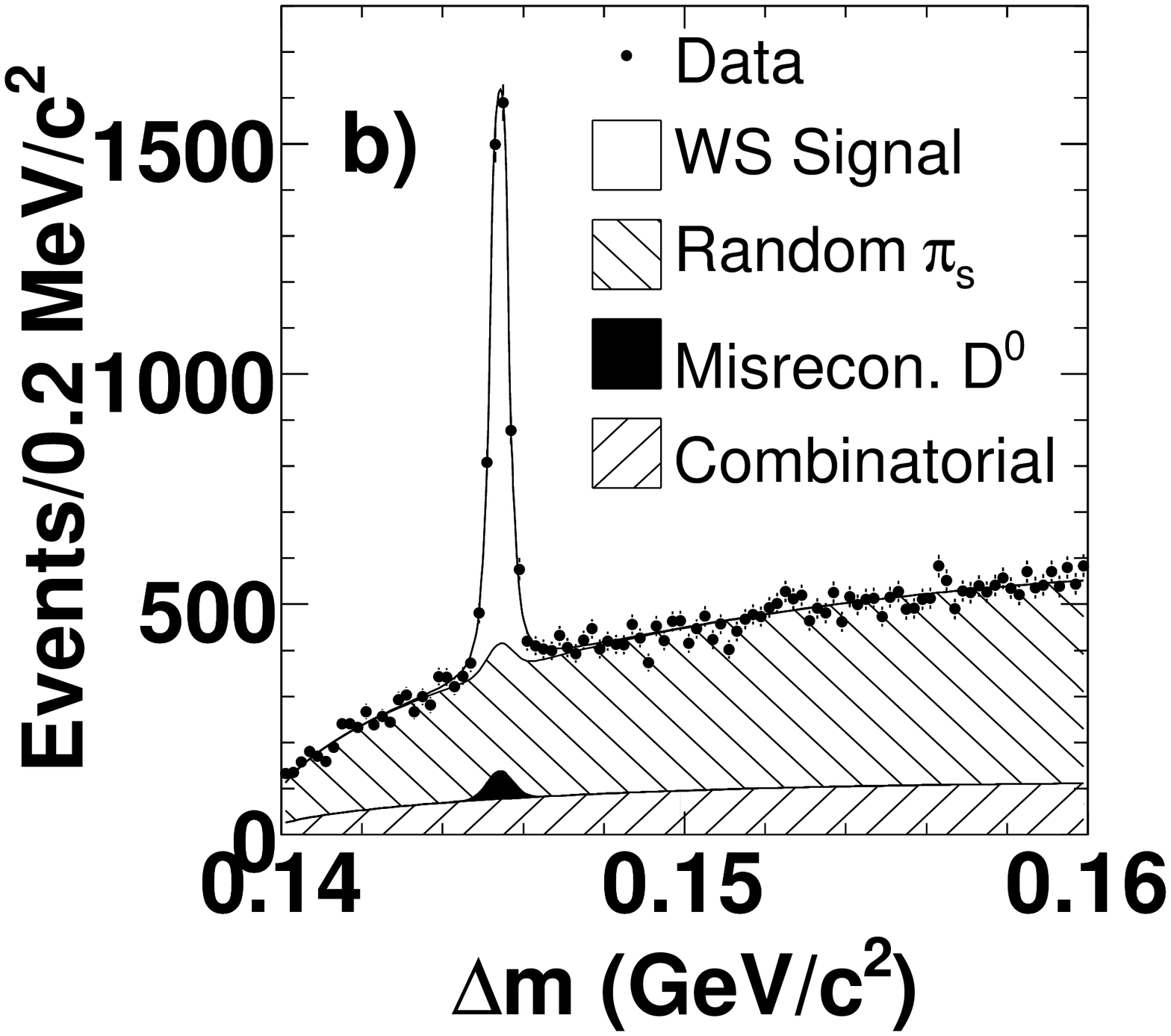}
  }
  \caption{a)  \mKpi for 
wrong-sign (WS) candidates with $0.1445<\dm<0.1465\gevcc$, and b)  \dm for WS candidates
with $1.843<\mKpi<1.883\gevcc$. The fitted PDFs are overlaid. 
}
  \label{fig:r18Data_MdMfit}
  \smallskip
\end{figure}

We fit the RS and WS data samples simultaneously with shape parameters
describing the signal and random $\pisoft^+$ components shared between
the two data samples. We find $1,141,500\pm
1,200$ RS signal events and $4,030\pm 90$ WS signal events. The
dominant background component is the random $\pisoft^+$ background.
Projections of the WS data and fit are shown in Fig.~\ref{fig:r18Data_MdMfit}.

The measured proper-time distribution for the RS signal is described by an
exponential function convolved with a resolution function whose
parameters are determined by the fit to the data. The resolution function
is the sum of three Gaussians with widths
proportional to the estimated event-by-event proper-time uncertainty
\terr. The random $\pisoft^+$ background is described by the same proper-time
distribution as signal events, since the slow pion has little weight
in the vertex fit. The proper-time distribution of the combinatorial
background is described by a sum of two Gaussians, one of which has a
power-law tail to account for a small long-lived component.  The
combinatorial background and real \Dz decays have different
\terr distributions, as determined from data using a background-subtraction 
technique based on the fit to \mKpi and \dm.

The fit to the RS proper-time distribution is performed over all
events in the full \mKpi and \dm region. The PDFs for signal and
background in \mKpi and \dm are used in the proper-time fit with all
parameters fixed to their previously determined values.  The fitted
\Dz lifetime is found to be consistent with the world-average
lifetime \cite{PDG2006}.

\begin{figure}[phtb]
  \centering
  \centerline{%
    \includegraphics[width=0.9\linewidth, clip=]{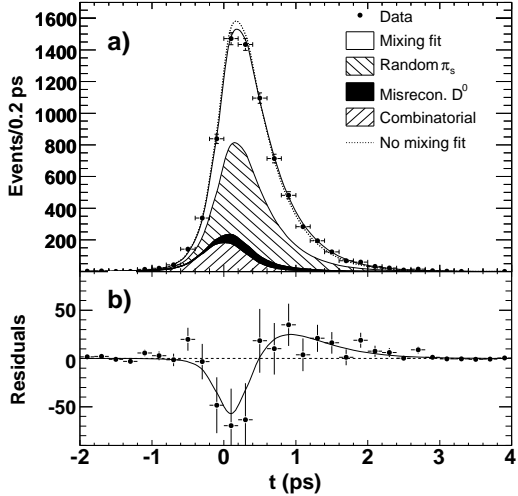}
  }
\caption{a) Projections of the 
proper-time distribution of combined \Dz and \Dzb 
WS candidates and fit result integrated over the signal region
$1.843<\mKpi<1.883\gevcc$ and $0.1445<\dm<0.1465\gevcc$.
The result of the fit allowing (not allowing) mixing 
but not \CP violation 
is overlaid as a solid (dashed) curve.
b) The points represent the difference 
between the data and the no-mixing fit. The solid curve
shows the difference between fits with and without mixing.}
\label{fig:histTimeBiasWSR18Data}
\end{figure}

The measured proper-time distribution for the WS signal is modeled by
Eq.~(\ref{eq:Tws}) convolved with the resolution function determined in
the RS proper-time fit. The random $\pisoft^+$ and misreconstructed
\Dz backgrounds are described by the RS signal proper-time
distribution since they are real \Dz decays. 
The proper-time distribution for WS data is shown in
Fig.~\ref{fig:histTimeBiasWSR18Data}. The fit results with and without 
mixing are shown as the overlaid curves.

\begin{figure}[phtb]
  \centering
  \centerline{%
    \includegraphics[width=0.95\linewidth, clip=]{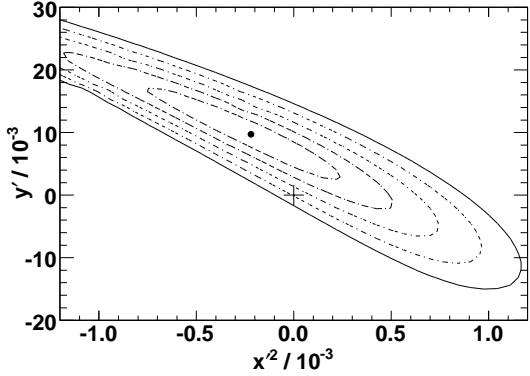}
}
\caption{The central value (point) and confidence-level (CL) contours for 
$1-\mbox{CL}=0.317\ (1\sigma)$, $4.55\times10^{-2}\ (2\sigma)$, 
$2.70\times10^{-3}\ (3\sigma)$, $6.33\times10^{-5}\ (4\sigma)$ and
$5.73\times10^{-7}\ (5\sigma)$, calculated from the change in the value
of $-2\ln{\cal L}$ compared with its value at the minimum.
Systematic uncertainties are included. The no-mixing point is shown 
as a plus sign~($+$).}
\label{fig:CPContour}
\end{figure}

The fit with mixing provides a substantially better description of the data
than the fit with no mixing.
The significance of the mixing signal is evaluated based on
the change in negative log likelihood with respect to the minimum.
Figure~\ref{fig:CPContour} shows confidence-level (CL) contours
calculated from the change in log likelihood ($-2\Delta\ln{\cal L}$) in two
dimensions (\xPrimeSq and \yPrime) with systematic uncertainties
included.  The likelihood maximum is at the unphysical value of
$\xPrimeSq=-2.2\times10^{-4}$ and $\yPrime = 9.7 \times 10^{-3}$. The
value of $-2\Delta\ln{\cal L}$ at the most likely point in the
physically allowed region ($\xPrimeSq=0$ and $\yPrime=6.4 \times
10^{-3}$) is $0.7$~units.  The value of
$-2\Delta\ln{\cal L}$ for no-mixing is $23.9$~units.
Including the systematic uncertainties, this corresponds to a
significance equivalent to 3.9~standard deviations
($1-\mbox{CL}=1\times10^{-4}$) and thus constitutes evidence for
mixing. The fitted values of the mixing parameters and \Rdcs are
listed in Table~\ref{tab:results}.  The correlation coefficient
between the \xPrimeSq and \yPrime parameters is $-0.94$.

\begin{table}[thb]
  \caption{Results from the different fits.
  The first uncertainty listed is statistical and the second systematic.}
  \label{tab:results}
  \centering\small

    \begin{tabular}{lcr@{~$\pm$}r@{~$\pm$}r}
\hline
     Fit type & Parameter & \multicolumn{3}{c}{Fit Results ($/10^{-3}$)}  \\
    \hline
    No \CP viol. or mixing & $\Rdcs$ & $3.53 $ & $ 0.08 $ & $ 0.04$\\
    \hline
    \multirow{3}{1.7cm}{No \CP\\ violation}
    &  $\Rdcs$        & $3.03$ & $0.16$ & $ 0.10$ \\
    &  $\xPrimeSq$  & $-0.22$ & $0.30$ & $ 0.21$   \\
    &  $\yPrime$    & $9.7$ & $4.4$ & $ 3.1$      \\
    \hline
    \multirow{5}{1.7cm}{\CP\\ violation \\ allowed}
    & $\Rdcs$     & $3.03$ &$0.16$ & $0.10$  \\
    & $\AD$       & $-21$ & $52$ & $15$  \\
    & $\xPrimePSq$ & $-0.24 $ & $ 0.43 $ & $ 0.30 $\\
    & $\yPrimeP$   & $ 9.8  $ & $ 6.4  $ & $ 4.5  $\\
    & $\xPrimeMSq$ & $-0.20 $ & $ 0.41 $ & $ 0.29 $\\
    & $\yPrimeM$   & $ 9.6  $ & $ 6.1  $ & $ 4.3  $\\
\hline
  \end{tabular}
\end{table}

\par
Allowing for the possibility of \CP violation, we calculate the values
of $\Rdcs = \sqrt{\Rdcs^+\Rdcs^-}$ and $\AD = (\Rdcs^{+} -
\Rdcs^{-})/(\Rdcs^{+} + \Rdcs^{-})$ listed in Table~\ref{tab:results},
from the fitted $\Rdcs^{\pm}$ values.  The best fit points
$(\xPrimePmSq,\yPrimePm)$ shown in Table~\ref{tab:results} are more
than three standard deviations away from the no-mixing hypothesis. The shapes of the
$(\xPrimePmSq,\yPrimePm)$ CL contours are similar to those shown
in Fig.~\ref{fig:CPContour}. All cross checks indicate that the close
agreement between the separate \Dz and \Dzb fit results is coincidental.

\par
\begin{figure}[phtb]
  \centering
  \centerline{%
    \includegraphics[width=0.9\linewidth, clip=]{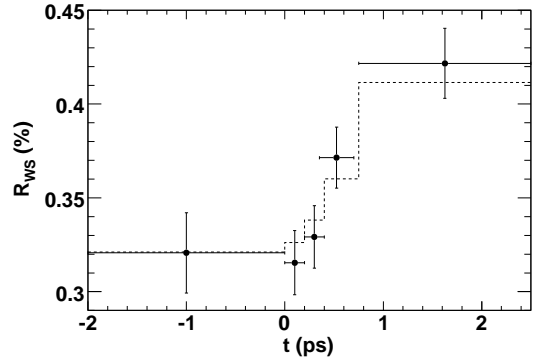}
  }
\caption{The WS branching fractions from independent \mdm\ fits to
slices in measured proper time (points). 
The dashed line shows the expected wrong-sign rate
as determined from the mixing fit shown in Fig.~\ref{fig:histTimeBiasWSR18Data}.
The $\chi^2$ with respect to expectation from the mixing fit is 1.5;
for the no-mixing hypothesis (a constant WS rate), the $\chi^2$ is 24.0.}
\label{fig:RwsTimeBins}
\end{figure}

As a cross-check of the mixing signal, we perform independent 
\mdm\ fits with no shared parameters for intervals in proper time selected
to have approximately equal numbers of RS candidates.
The fitted WS branching fractions are shown in
Fig.~\ref{fig:RwsTimeBins} and are seen to increase with time.  The
slope is consistent with the measured mixing parameters and
inconsistent with the no-mixing hypothesis.
 
We validated the fitting procedure on simulated data samples using both MC
samples with the full detector simulation and large parametrized MC
samples.  In all cases we found the fit to be unbiased. As a further
cross-check, we performed a fit to the RS data proper-time distribution
allowing for mixing in the signal component; the fitted values of the
mixing parameters are consistent with no mixing.


\par
In evaluating systematic uncertainties in \Rdcs and the mixing parameters
we considered variations in the fit model and in the selection criteria.
We also considered alternative forms of the \mKpi, \dm, proper time, and \terr
PDFs.  We varied the $t$ and \terr requirements.
In addition, we considered 
variations that keep or reject all
\Dstp\ candidates sharing tracks with other candidates. 
\par
For each source of systematic error, we compute the significance
$\signif_i^2=2t\big[\ln\like\xPSQyP-\ln\like(\xPrimeSq_i, \yPrime_i)\big]/2.3$, where 
\xPSQyP are the parameters obtained from the standard fit, 
$(\xPrimeSq_i, \yPrime_i)$ the parameters from the fit including the $i^{th}$ systematic variation,
and \like\ the likelihood of the standard fit. 
The factor 2.3 is the 68\% confidence level for 2 degrees of freedom. 
To estimate the significance of our results in \xPSQyP, 
we reduce $-2\Delta\ln{\cal L}$ by a factor of $1+\Sigma s_i^2=1.3$
to account for systematic errors. The largest contribution to this factor, $0.06$, is due to uncertainty in modeling
the long decay time component from other $D$ decays in the signal region. The second largest 
component, $0.05$, is due to the presence of a non-zero mean in the proper time signal resolution
PDF.
The mean value is determined in the RS proper time fit to be $3.6\fs$ and is due to
small misalignments in the detector.
The
error of $15\times 10^{-3}$ on \AD is primarily due to uncertainties in modeling the differences
between \Kp and \Km absorption in the detector.

\par
In conclusion we summarize the \babar\  evidence
for $\Dz$-$\Dzb$ mixing.
Our result is inconsistent with the no-mixing hypothesis 
at a significance of 3.9 standard deviations. 
We measure $\yPrime=[9.7 \pm 4.4 \hbox{ (stat.)}\pm 3.1 \hbox{ (syst.)}] \times 10^{-3}$,
while \xPrimeSq is consistent with zero. We find no evidence for \CP
violation and measure \Rdcs to be $[0.303\pm0.016\hbox{ (stat.)}\pm 
0.010\hbox{ (syst.)}]\%$.  
The result is consistent with
Standard Model estimates for mixing.


\end{document}

%% file: note1706symbols.tex
\def\Rdcs       {\ensuremath{R_{\rm D}}\xspace} 

\def\xPrimeSq   {\ensuremath{{x^{\prime}}^2}\xspace}

\def\yPrime     {\ensuremath{y^{\prime}}\xspace}
\def\yPrimeP    {\ensuremath{y^{\prime+}}\xspace}
\def\yPrimeM    {\ensuremath{y^{\prime-}}\xspace}
\def\xPrimePmSq {\ensuremath{{x^{\prime}}^{2\pm}}\xspace}

\def\yPrimePm   {\ensuremath{y^{\prime\pm}}\xspace}

\def\xPrimePSq  {\ensuremath{{x^{\prime}}^{2+}}\xspace}
\def\xPrimeMSq  {\ensuremath{{x^{\prime}}^{2-}}\xspace}
\def\xPrimePmSq  {\ensuremath{{x^{\prime}}^{2\pm}}\xspace}
\def\xPSQyP     {\ensuremath{(\xPrimeSq,\,\yPrime)}\xspace}

\def\AD         {\ensuremath{A_{\rm D}}\xspace}

\def\pisoft     {\ensuremath{\pi_{\rm s}}\xspace}

\def\Dztokpi    {\ensuremath{\Dz \to K^{-}\pi^{+}}\xspace}

\def\Dzbtokpi   {\ensuremath{\Dzb \to K^{+}\pi^{-}}\xspace}
\def\DztokpiWS  {\ensuremath{\Dz \to K^{+}\pi^{-}}\xspace}

\def\Kmpip      {\ensuremath{K^{-}\pi^{+}}\xspace}
\def\Kmp        {\ensuremath{K^{\mp}}\xspace}

\def\KpKm       {\ensuremath{K^{+}K^{-}}\xspace}

\def\pimp       {\ensuremath{\pi^{\mp}}\xspace}
\def\pipm       {\ensuremath{\pi^{\pm}}\xspace}

\newcommand{\kevcc}{\ensuremath{{\mathrm{\,Ke\kern -0.1em V\!/}c^2}}\xspace}
\def\mKpipis    {\ensuremath{m_{K\pi\slowpi}}\xspace}
\def\mKpi       {\ensuremath{m_{K\pi}}\xspace}
\def\dm         {\ensuremath{\Delta m}\xspace}
\def\t          {\ensuremath{t}}
\def\terr       {\ensuremath{\sigma_{\t}}\xspace}
\def\chisq       {\ensuremath{\chi^2}}
\def\Pchisq      {\ensuremath{P(\chi^2)}}

\def\Dstp        {\ensuremath{D^{*+}}}

\def\slowpi      {\ensuremath{\pi_s}}

\def\like   {\ensuremath{{\cal L}}}
\def\signif {\ensuremath{{s}}}

\def\mdmnospace            {\ensuremath{\{\mKpi,\,\dm\}}}
\def\mdm            {\mdmnospace\xspace}

\def\Pbar    {\ensuremath{\kern 0.2em\overline{\kern -0.2em P}{}}\xspace}

\def\Dm      {\ensuremath{\Delta m}\xspace}

\catcode`\@=11\relax
\newskip\dkwidth
\def\dk{%
   \dkwidth=2em plus 0.5 em minus 0.25 em\relax
   {\m@th\mathord{%
   \hbox{%
      \kern 0.3em
      \raise 0.6ex%
      \hbox{%
         \vrule width 0.25pt height 0.5\dkwidth depth0pt}%
      \kern-1.2pt%
      \hbox to 1.1\dkwidth{%
         \rightarrowfill}%
      \kern0.4em}}%
   }%
}%
\def\rightarrowfill{$\m@th\mathord-\mkern-10mu%
  \cleaders\hbox{$\mkern-2mu\mathord-\mkern-2mu$}\hfill
  \mkern-6mu\mathord\rightarrow$}
\catcode`\@=12\relax

%% file: D0mixing.bbl
\begin{thebibliography}{9}
\bibitem{footnote} The use of charge-conjugate modes is implied unless otherwise
noted.
\end{thebibliography}
